# Chapter 13

# Beam Instrumentation and Long-Range Beam–Beam Compensation

E. Bravin, B. Dehning, R. Jones\*, T. Lefevre and H. Schmickler

CERN, Accelerator & Technology Sector, Geneva, Switzerland.

## 13  Beam instrumentation and long-range beam–beam compensation

### 13.1  Introduction

The extensive array of beam instrumentation with which the LHC is equipped has played a major role in its commissioning, rapid intensity ramp-up, and safe and reliable operation. In addition to all of these existing diagnostics, the HL-LHC brings a number of new challenges in terms of instrumentation that are currently being addressed.

The beam loss system, designed to protect the LHC from losses that could cause damage or quench a superconducting magnet, will need a significant upgrade in order to be able to cope with the new demands of the HL-LHC. In particular, cryogenic beam loss monitors are under investigation for deployment in the new inner triplet magnets to distinguish between collision debris and primary beam losses. Radiation-tolerant integrated circuits are also under development to allow the front-end electronics to sit much closer to the detector, so minimizing the cable length required and reducing the influence of noise.

The use of crab cavities and possible use of long-range beam–beam compensators and hollow-electron lenses also implies new instrumentation in order to allow for optimization of their performance. Several additional diagnostic systems will therefore be considered. Very high bandwidth pick-ups and a streak camera installation to perform intra-bunch transverse position measurements are being investigated, along with new techniques for transverse beam size measurements such as a beam gas vertex detector.

An upgrade to several existing systems is also envisaged, including the beam position measurement system in the interaction regions and the addition of a halo measurement capability to synchrotron light diagnostics.

### 13.2  Beam loss measurement

Monitoring of beam losses is essential for the safe and reliable operation of the LHC. The beam loss monitoring (BLM) system provides knowledge of the location and intensity of such losses, allowing an estimation to be made of the energy dissipated in the equipment along the accelerator. The information is used for machine protection, to qualify the collimation hierarchy, to optimize beam conditions, and to track the radiation dose to which equipment has been exposed. This is currently done using nearly 4000 ionization monitors distributed around the machine and located at all probable loss locations, with the majority mounted on the outside of the quadrupole magnets, including those in the inner triplet regions. Around one-third of the arc monitors have recently been relocated in order to optimize the system for protection against fast beam losses believed to be caused by dust particles falling into the vacuum pipe. While the existing system is believed to meet the needs of the HL-LHC for the arcs, this will no longer be the case for the high luminosity interaction points.

---

\* Corresponding author: Rhodri.Jones@cern.ch

In the HL-LHC high luminosity insertions the magnets will be subjected to a greatly enhanced continuous radiation level due to the increase in collision debris resulting from the higher luminosity. With the presently installed configuration of ionization chambers in this region the additional signal from any dangerous accidental losses would be completely masked by that coming from collision debris. This is a critical issue for LHC machine protection and therefore R&D has started to investigate possible options for placing radiation detectors inside the cryostat of the triplet magnets as close as possible to the superconducting coils. The dose measured by such detectors would then correspond much more precisely to the dose deposited in the coils, allowing the system to be used once again to prevent a quench or damage.

The quench level signals estimated for 7 TeV running are, for some detectors, very close to the noise level of the acquisition system. This is mainly determined by the length of cable required to bring the signal from the radiation hard detector to the more radiation-sensitive front-end electronics. Although qualified for use in the low radiation environments of the LHC arcs, the current electronics cannot be located close to the detectors in the higher radiation insertion regions. Development has thus started to implement these electronics in a radiation-hard application-specific integrated circuit (ASIC).

### 13.2.1 Beam loss monitors for the HL-LHC triplet magnets

Three detectors are currently under investigation as candidates for operation at cryogenic temperatures inside the cryostat of the triplet magnets [1]:

- single crystal chemical vapour deposition (CVD) diamond with a thickness of 500 μm, an active area of 22 $mm^2$, and gold as the electrode material;

- p+–n–n+ silicon wafers with a thickness of 280 μm, an active area of 23 $mm^2$ and aluminum as the electrode material;

- liquid helium ionization chambers.

Experiments have already been performed to observe the behaviour of such detectors in a cryogenic environment and on the radiation effects at such temperatures upon silicon and single crystal diamond. Irradiation at up to 2 MGy ($0.8 \times 10^{14}$ protons/$cm^2$) showed degradation in the charge collection efficiency for CVD diamond by a factor of 15 and for Si by a factor of 25 (see Figure 13-1). The major downside of silicon compared to diamond, its much higher leakage current when irradiated, has been observed to disappear at liquid helium temperatures, with the leakage current remaining below 100 pA at 400 V, even under forward bias for an irradiated diode. Further experiments combining irradiation with cryogenic temperatures will be necessary to optimize the final detector design. These experiments will be accompanied by tests of detectors mounted inside the cryostats of existing LHC magnets with the aim of gaining experience with the long-term performance of such detectors under operational conditions. The technology chosen will need to able to withstand irradiation up to 20 MGy at 4.5K.

Up to six detectors will be installed inside the cold mass of each main triplet quadrupole magnet, which leads to a baseline procurement of 100 detectors (96 installed and four spares). If the option of equipping the 11 T dipole and all the spare triplet magnet assemblies is also taken into account, then a total procurement of 150 detectors would be required. They are foreseen to be housed in existing holes in the iron yoke within the cold mass of the magnet. Each detector will be equipped with a single semi-rigid coaxial cable that will provide the necessary high voltage (up to 1000 V) and extract the loss signal from each detector. Feedthroughs allowing a total of six coaxial cable connections will need to be integrated into each of the main triplet quadrupole cryostats with the location of the feedthrough chosen so as to minimize the cable length required.

As part of the machine protection system these components need to be highly reliable and maintenance-free. In the event that some of these monitors stop functioning the existing external BLM system should still provide adequate protection against damage due to excessive beam loss, but will probably not be able to distinguish quench provoking losses from the experimental background.

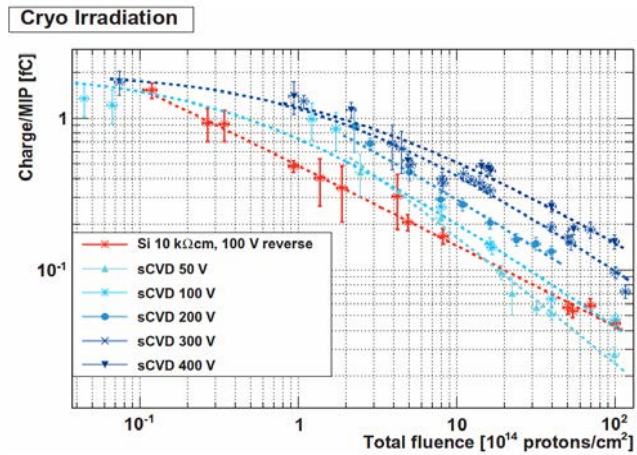

Figure 13-1: Charge collection efficiency for silicon and diamond detectors with increasing radiation fluence in a cryogenic environment.

13.2.2    A rad-tolerant application-specific integrated circuit for the HL-LHC beam loss monitoring system

The current front-end electronics for the LHC BLM system, while providing a 40 μs integration time, is limited in the dynamic range it can handle and is only radiation tolerant up to ~500 Gy. The latter implies the use of long cables in the higher radiation LSS regions, which further limits the dynamic range and in some cases brings the noise floor close to the quench level signal at 7 TeV. Instead of the discrete component currently used, an optimized ASIC is therefore under development. This is still based on the current-to-frequency conversion used in the existing system, but is packaged in a compact, radiation-tolerant form with an increased dynamic range. The technique employed allows the digitization of bipolar charge over a 120 dB dynamic range (corresponding to an electric charge range of 40 fC–42 nC) with a 40 μs integration time and a conversion reference provided by an adjustable, temperature-compensated current reference [2].

Figure 13-2 shows the block diagram of the integrated circuit. It is composed of a bipolar, fully-differential integrator that converts the charge received from the detector into a voltage input for a synchronous comparator system. A three-level digital-to-analog converter (DAC) drives the discharge current for the integrator and is connected in a feedback loop to the comparator output. The first logic block is used to select the gain in the integrator, the current step in the DAC, and the threshold in the comparators, while a second logic block encodes the output signal from the comparators. The results of both of these are used by a third block to assemble the final, correctly weighted, output word.

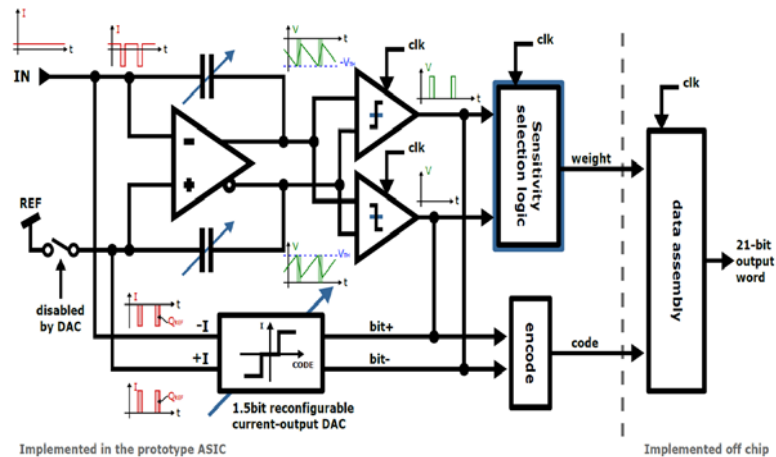

Figure 13-2: Schematic representation of the ASIC implementation currently under development

The prototype ASIC is designed with commercial CMOS technology and has two analog-to-digital (A/D) channels and a sensitivity selection logic that can be disabled to implement the circuitry externally. This strategy has been useful for testing the device and improving the algorithm. Its measured characteristics are listed in Table 13-1.

Table 13-1: Characteristics of the prototype HL-LHC BLM system ASIC

| Parameter | Value |
| --- | --- |
| **A/D converter** | |
| Integration time | 40 μs |
| Input current range | −1.05–1.05 mA |
| Input charge range | −42–42 nC |
| Offset | <40 aC at 40 μs integration, <1 pA |
| Default least significant bit step | 50 fC ±20%, adjustable |
| Dynamic range | 120 dB |
| Linearity error | <±5% |
| Peak signal/noise ratio | 53 dB |
| SFDR at 999 Hz, 1 mA | 50 dB |
| Total ionizing dose | 10 Mrad (Si) |
| Supply voltage | 2.5 V |
| Clock | 12.8 MHz |
| Power | 40 mW |
| **Reference charge** | |
| Drift with TID | 3% at 10 Mrad (Si) |
| Drift with temperature | $<600 \times 10^6$/°C |

The measured linearity is limited by transistor matching imperfections in the DAC, introducing an error at the transition between the sensitivities. However, overall the error is less than 5% and well inside specification (<10%).

Total ionizing dose (TID) effects on the ASIC have been investigated using an X-ray beam with 20 keV peak energy. The characteristics of the device were measured up to 100 kGy (Si), followed by a one-week annealing cycle at 100°C. From the beginning to the end of the irradiation cycles, the functionality was always preserved, with the conversion offset remaining below 1 least significant bit (LSB) and the value of the full-scale charge drifting by less than 3%. Development will now continue to address the issues found using the prototype and to implement more advanced logic blocks within the ASIC.

It is foreseen that a total of some 300 detectors, mainly located in the LSS regions, will be equipped with such a front-end ASIC.

**13.3  Beam position monitoring**

With its 1070 monitors, the LHC beam position monitor (BPM) system is the largest BPM system in the world [3]. Based on the wide band time normalizer (WBTN) principle, it provides bunch-by-bunch beam position over a wide dynamic range (~50 dB). Despite its size and complexity (3820 electronics cards in the accelerator tunnel and 1070 digital post-processing cards in surface buildings) the performance of the system during the last three years has been excellent, with greater than 97% overall availability.

13.3.1   Current performance and limitations

The position resolution of the LHC arc BPMs has been measured to be better than 150 μm when measuring a single bunch on a single turn and better than 10 μm for the average position of all bunches. The main limitation on the accuracy of the BPM system is linked to temperature-dependent effects in the acquisition electronics, which can generate offsets of up to a millimetre over a timescale of hours. On-line compensation was introduced to limit this effect during operation, and temperature-controlled racks are currently being installed with the hope of eliminating this limitation from Run 2 start-up in 2015.

The non-linearity of the BPMs located near the interaction points has also proven to be problematic, in particular for accurate measurements during the beta squeeze and during machine development periods. A new correction algorithm has therefore been developed, based on exhaustive electromagnetic simulations, with the aim of bringing down the residual error to below 20 μm over most of the useable BPM area [4]. Developed to be able to distinguish between the positions of two counter-propagating beams in the same beam pipe, these BPMs also suffer from non-optimal decoupling between the beams, which is something that will need to be addressed for the HL-LHC.

### 13.3.2 A high resolution orbit measurement system for the HL-LHC

Originally developed to process signals from BPM buttons embedded in LHC collimator jaws, orbit measurement using a compensated diode detector scheme [5] has already been demonstrated to be simple and robust, and to provide a position resolution down to the nanometre level. A comparison of the orbit measured on a single BPM during a van der Meer scan by the current orbit system and the new diode orbit system is presented in Figure 13-3, where the resolution of the new system can be seen to be over 50 times better. All new LHC collimators will have BPMs using this acquisition system installed with them, with plans to also equip the BPMs in all four LHC interaction regions. It is important to note, however, that the new system does not provide the bunch-by-bunch measurement capability of the existing system.

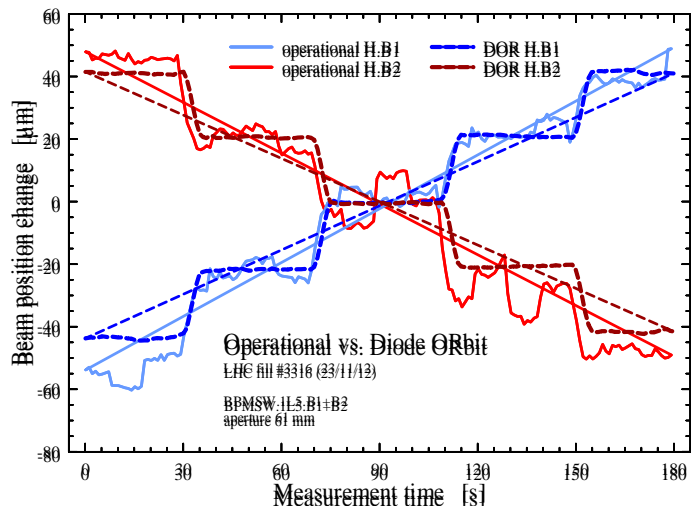

Figure 13-3: Comparison of the new LHC orbit system electronics with the existing system during a van der Meer scan.

At the start of the HL-LHC era the existing BPM system will have been operational for over 15 years, using components that are over 20 years old. It is therefore likely that a completely new system will need to be installed for HL-LHC operation. One candidate would be to extend the new diode orbit system to the whole machine for accurate global orbit measurements, and complement this with a system capable of providing the high-resolution bunch-by-bunch, turn-by-turn measurements required in particular for optics studies and the many other accelerator physics experiments that will be needed to understand and optimize HL-LHC performance.

With the higher bunch intensities foreseen, the dynamic range of the BPM system for the HL-LHC will need to be increased accordingly. The present system implements two sensitivity ranges, optimized for pilot and nominal bunch intensities. Issues have been observed in the first three years of operation with BPMs providing large errors when reaching the limit of their dynamic range. For the interlock BPMs located in P6, this can trigger false beam dumps, which clearly has an impact on machine availability. Although improvements have already been made to this interlock system for Run 2, any consolidation of the LHC BPM system should include developing dedicated electronics for this system, optimized for both high reliability and availability.

### 13.3.3 High directivity strip-line pick-ups for the HL-LHC insertion regions

In the BPMs close to the interaction regions, the two beams circulate in the same vacuum chamber. Directional strip-line pick-ups are therefore used to distinguish between the positions of both beams. When the two beams pass through the BPM at nearly the same time, the two signals interfere due to the limited directivity of the strip-line that, in the present design, only gives a factor of 10 isolation between the wanted signal and that coming from the other beam. This effect can be minimized by installing the BPMs at a location where the two beams do not overlap temporally. This is a constraint included in both the present and future layout, but which cannot be satisfied for all BPM locations. The ideal longitudinal location corresponds to $(1.87 + N \times 3.743)$ m from the IP where $N$ is an integer. Any deviation from this will diminish the possibility of the system distinguishing one beam from the other.

For the HL-LHC BPMs in front of the Q2a and Q3 magnets and the triplet corrector magnet package, there is the additional constraint that tungsten shielding is required in the cold bore to minimize the heat deposition in these magnets. A mechanical re-design coupled with extensive electromagnetic simulations is therefore necessary to optimize the directivity under these constraints.

As part of the beam position system of the LHC these components need to be highly reliable and maintenance-free, while the system should be able to measure the beam position for each beam with a resolution of 1 μm and a medium term (fill to fill) reproducibility of 10 μm.

#### 13.3.3.1 Cold stripline beam position monitors

The HL-LHC high luminosity insertion regions will be equipped with two types of cold stripline BPMs, measuring simultaneously the position of both beams in both planes.

The BPMs located in the interconnect in front of the Q2b and Q3 magnets and the triplet corrector magnet package will be cold stripline BPMs, rotated by 45° to allow the insertion of tungsten shielding in the median planes of both horizontal and vertical axes.

The BPMs located in front of Q2a and before and after the D1 magnet will be cold stripline BPMs with orthogonally positioned electrodes and without tungsten shielding.

The signal from all of these BPMs will be extracted using eight semi-rigid, radiation-resistant coaxial cables per BPM. Two feedthroughs with four coaxial cable connections will be integrated into the Q2a, Q2b, and Q3 cryostats and into the cryostat of the triplet corrector magnet package, with four such feedthroughs integrated into the D1 cryostat. The outputs on these feedthroughs will be connected to standard ½″ coaxial cables taking the signal to the electronics in the UA/UJ.

A total of 12 stripline BPMs of each type will be installed, with three spares foreseen per type of BPM assembly.

#### 13.3.3.2 Warm stripline beam position monitors

The beam position monitor in front of Q1a will be a warm stripline BPM, simultaneously measuring the position of both beams in both planes. The signal will be extracted using eight semi-rigid, radiation-resistant coaxial cables, to a patch panel located in an area of lower radiation on the tunnel wall, where they will connect to eight standard ½″ coaxial cables taking the signal to the electronics in the UA/UJ.

A total of four such BPMs will be installed with two spares foreseen for this type of BPM assembly.

### 13.3.4 Collimator beam position monitors

All next-generation collimators in the LHC will have button electrodes embedded in their jaws for on-line measurement of the jaw-to-beam position [4]. This is expected to provide a fast and direct way of positioning the collimator jaws and subsequently allow constant verification of the beam position at the collimator location, improving the reliability of the collimation system as a whole. The design of such a BPM was intensively simulated using both electromagnetic (EM) and thermo-mechanical simulation codes. In order to provide the best accuracy, the BPM readings must be corrected for the nonlinearity coming from the varying geometry of the collimator jaws as they are closed and opened, for which a 2D polynomial correction has been obtained from EM simulations and qualified with beam tests using a prototype system installed in the CERN SPS.

The collimator BPM hardware, i.e. the button electrode located in the jaw, the cable connecting the electrode to the electrical feedthrough mounted on the vacuum enclosure, and the feedthrough itself have been chosen to withstand the radiation dose of 20 MGy expected during the lifetime of the collimator.

## 13.4 Beam profile measurements

The LHC is currently fitted with a host of beam size measurement systems used to determine beam emittance. These different monitors are required in order to overcome the specific limitation of each individual system. Wirescanners are used as the absolute calibration reference, but can only be operated with a low number of bunches in the machine due to intensity limitations linked to wire breakage. A cross-calibrated synchrotron light monitor is therefore used to provide beam size measurements, both average and bunch-by-bunch, during nominal operation. However, the small beam sizes achieved at 7 TeV, the multiple sources of synchrotron radiation (undulator, D3 edge radiation, and central D3 radiation), and the long optical path required to extract the light imply that the correction needed to extract an absolute value is of the same order of magnitude as the value itself. This requires an excellent knowledge of the error sources to obtain meaningful results. The third system installed is an ionization profile monitor, which is foreseen to provide beam size information for lead ions at injection, when there is insufficient synchrotron light. The monitor has also been used for protons, but suffers from significant space charge effects at energies above 2 TeV.

Whilst efforts are ongoing to improve the performance of all the above systems, alternative techniques to measure the transverse beam size and profile are also under study for the HL-LHC.

### 13.4.1 Fast wirescanners

The currently installed LHC linear wirescanners have a maximum scan speed of 1 ms$^{-1}$. This gives a limit on the total intensity that can be scanned at injection of ~$2.7 \times 10^{13}$ protons for an emittance of ~2 µm, or some 200 nominal bunches. Scanning at 20 ms$^{-1}$ would allow systematic, average beam size measurements to be performed on the full physics beam at injection. A new fast, rotational wirescanner concept is therefore being explored to reach such speeds whilst achieving an accuracy of 5 µm on the beam width determination, i.e. an error of 5% or less for nominal emittance beams.

The mechanics will be based on the design currently being developed for the injector complex as part of the LHC Injector Upgrade project [6]. It eliminates the need for mechanical bellows by placing all moveable parts of the rotational scanner in the beam vacuum. These bellows have a limited lifetime that, due to their intensive use in the LHC, can be reached within a matter of years. The use of a magnetically coupled motor without the need for moving vacuum parts should significantly increase the MTBF of the mechanical system. In addition, the current shower detection acquisition will be replaced with a diamond-based sensor connected to high dynamic range acquisition electronics, which should considerably improve the operational ease-of-use of these devices.

Plans to install two horizontal and two vertical scanners for Beam 1 next to the existing scanners in 5R4 and two horizontal and two vertical scanners for Beam 2 next to the existing scanners in 5L4 are foreseen. One spare horizontal and one spare vertical unit will also be manufactured.

### 13.4.2  A Beam gas vertex profile monitor (BGV)

The VELO detector of the LHCb experiment has shown how beam–gas interactions can be used to reconstruct the transverse beam profile of the circulating beams in the LHC [7]. The new concept under study is to see whether a simplified version of such a particle physics tracking detector can be used to monitor the beams throughout the LHC acceleration cycle. This concept has, until now, never been applied to the field of beam instrumentation, mainly because of the high level of data treatment required. However, the advantages compared to the standard beam profile measurement methods listed above are impressive: high resolution profile reconstruction, single bunch measurements in three dimensions, quasi non-destructive, no detector equipment required in the beam vacuum, and high radiation tolerance of the particle detectors and accompanying acquisition electronics.

Such a beam shape measurement technique is based on the reconstruction of beam–gas interaction vertices, where the charged particles produced in inelastic beam–gas interactions are detected with high-precision tracking detectors. Using the tracks left in the detectors, the vertex of the particle–gas interaction can be reconstructed so, with enough statistics, building up a complete 2D transverse beam profile (Figure 13-4). The longitudinal profile could also be reconstructed in this way if relative arrival time information is additionally acquired by the system, which is not currently planned.

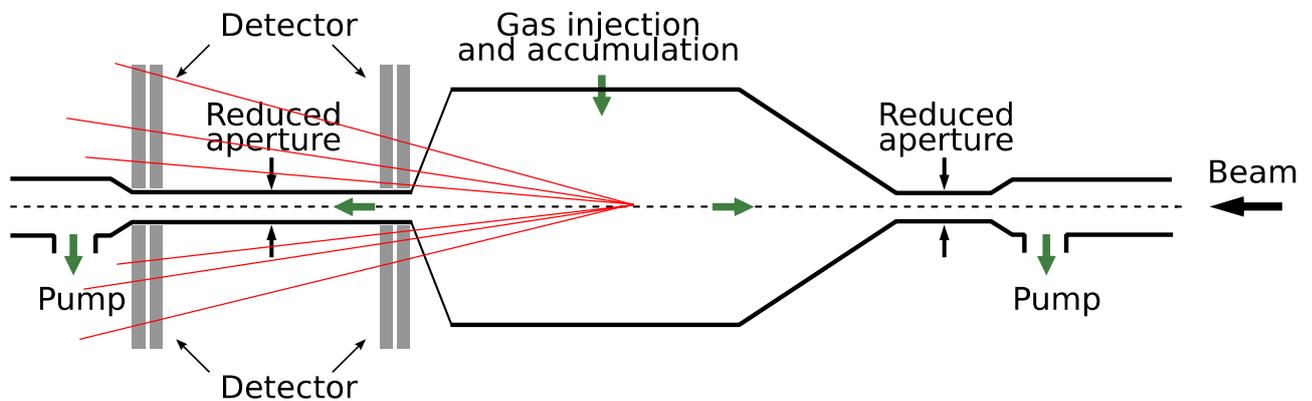

Figure 13-4: A sketch demonstrating the beam gas vertex measurement concept

Unlike LHCb, where the detector is placed very close to the beam and can therefore only be used during stable beams, the aim with the beam gas vertex profile monitor (BGV) detector is to design a robust instrument that can be used for beam size measurements throughout the LHC cycle. Its final specifications are to provide:

- transverse bunch size measurements with a 5% resolution within 1 minute;
- average transverse beam size measurements with an absolute accuracy of 2% within 1 minute.

The main subsystems are: a neon gas target at a pressure of $6 \times 10^{-8}$ mbar, a thin aluminium exit window, tracking detector based on scintillating fibre modules read out by silicon photomultipliers, hardware and software triggers, and a readout and data acquisition system based on that used for LHCb. As the tracking detector is external to the vacuum chamber, no movable parts are needed. The final design of the prototype is shown in Figure 13-5.

A proof-of-principle demonstrator is foreseen for installation on the lefthand side of LHC IP4 on Beam 2 during LS1, with a second system for Beam 1 installed on the righthand side of IP4 in LS2. A full upgrade of these systems, to include a third detector layer and upgraded electronics, may be required in LS3 to reach the full design goals outlined above. This upgrade is currently not part of the HL-LHC baseline.

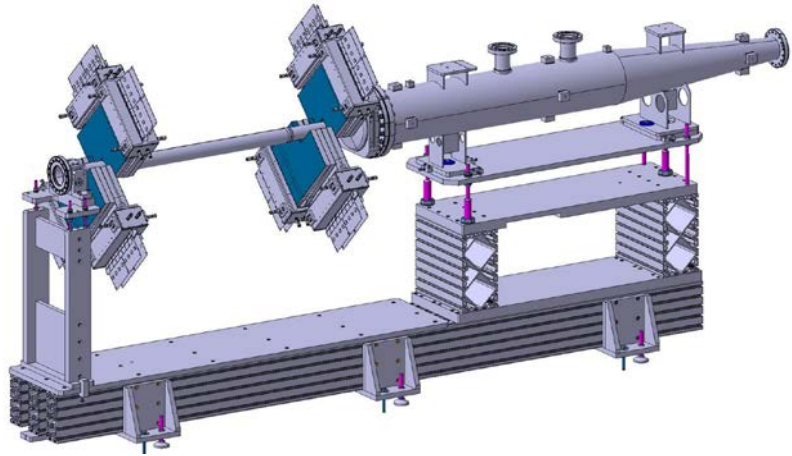

Figure 13-5: The demonstrator beam gas vertex detector installed in the LHC during LS1

## 13.5 Diagnostics for crab cavities

The crab cavities for the HL-LHC are proposed to counter the geometric reduction factor and so to enhance luminosity. These cavities will be installed around the high luminosity interaction points (IP1 and IP5) and used to create a transverse bunch rotation at the IP. The head and tail of each bunch is kicked in opposite directions by the crab cavities such that the incoming bunches will cross parallel to each other at the interaction point. These intra-bunch bumps are closed by crab cavities acting in the other direction on the outgoing side of the interaction region. If the bumps are not perfectly closed the head and tail of the bunch will follow different orbits along the ring. Monitors capable of measuring the closure of the head–tail bump and any head–tail rotation/oscillation outside of the interaction regions are therefore required.

### 13.5.1 Bunch shape monitoring using electromagnetic pick-ups

Electromagnetic monitors for intra-bunch diagnostics are already installed in the LHC. These so-called 'head–tail' monitors mainly provide information on instabilities and have a bandwidth of approximately 2 GHz. To go to a higher resolution within the bunch a bandwidth of 10 GHz or more is desirable. This will be important to better understand instabilities in the HL-LHC and to help with the tuning of the crab cavities, with several of these systems foreseen for installation around the interaction points. In addition to studies aimed at improving the existing electromagnetic pick-ups, which include optimization of the pick-up design and the testing of faster acquisition systems, pick-ups based on electro-optical crystals in combination with laser pulses are also being considered [8]. Such pick-ups have already demonstrated fast time responses in the picosecond range [9]. Developed mainly for linear accelerators, this technology is now also being considered for circular machines, with a design for a prototype to be tested on the CERN SPS recently initiated.

### 13.5.2 Bunch shape monitoring using streak cameras

The use of synchrotron light combined with a streak camera may be an easier alternative to electromagnetic or electro-optical pick-ups for high resolution temporal imaging. Using an optical system to re-image the synchrotron light at the entrance of a streak camera allows the transverse profile of the beam to be captured in one direction (X or Y) with a very fast time resolution (below the picosecond level). Only one transverse axis can be acquired with a given setup, while the other is used for the streaking. Using a sophisticated optical setup it is, however, possible to monitor both axes at the same time, as was performed in LEP [10].

Streak cameras can be used to observe a number of parameters simultaneously: bunch length, transverse profiles along the bunch, longitudinal coherent motion, head–tail motion, etc. The main limitation of the streak camera is the repetition rate of the acquisition, typically <50 Hz, and the limited length of the recorded sample, given by the CCD size. Double scan streak cameras exist that allow an increase in the record length. By using a CCD with 1000 × 1000 pixels working at 50 Hz and adjusting the optical magnification and scan speed such that the image of each bunch covers an area of about 100 × 100 pixels, it is possible to record a maximum of 100 bunch images per 20 ms, i.e. 5000 bunches/s. This is clearly just an optimistic upper limit with other factors likely to reduce this value.

The longitudinal resolution of around 50 ps required for the HL-LHC is rather easy to achieve using streak cameras, where measurements down to the sub-picosecond range are now possible. In terms of transverse resolution two distinctions have to be made.

- Measurement of the beam width is affected by diffraction due to the large relativistic gamma of the beam, with the diffraction disk of the same order as the beam size. This will significantly reduce the resolution of such measurements.

- The centroid motion (i.e. the centre of gravity) is not directly affected by diffraction, which produces a symmetrical blur, and therefore the resolution for this type of measurement will be much better.

For determining crab cavity non-closure only the average position along the bunch is of importance, and not changes to the beam size. The streak camera should therefore be able to achieve a resolution of a few percent of the beam sigma for this measurement.

Streak cameras are expensive and delicate devices not designed for the harsh environment inside an accelerator. Radiation dose studies are therefore required in order to verify if a streak camera can be installed directly in the tunnel or if, which seems more likely, it has to be housed in a dedicated, shielded, hutch. The latter would imply an optical line to transport the synchrotron light from the machine to the camera, something for which an integration study will be initiated.

Another point to consider is the synchrotron light source. Currently, two synchrotron light telescopes are installed in the LHC, one per beam. These telescopes already share their light amongst three different instruments, the synchrotron light monitor, the abort gap monitor, and the longitudinal density monitor, and in the future will also have to accommodate halo diagnostics (see below). It will therefore be difficult to integrate yet another optical beam line for the streak camera. The installation of additional light extraction mirrors will therefore be necessary to provide the light for the streak cameras. Since the crab cavities are only needed at high energy, dipole magnets can be used as the source of the visible synchrotron radiation for the streak cameras, with no need for the installation of additional undulators that are only required at injection energy, where the dipole radiation is in the infra-red. The best location and corresponding light extraction system for such an additional synchrotron radiation source is currently under study.

### 13.6 Halo diagnostics

Population of the beam 'halo', i.e. particles in between the beam core and the limits set by the primary collimators, can lead to important loss spikes through orbit jitter at the collimator locations. Measurement of the beam halo distribution is therefore important for understanding and controlling this mechanism. Such measurements are also important to determine the effectiveness of equipment that influences the beam halo, such as hollow electron lenses or long-range beam–beam compensators. Moreover, in the HL-LHC, any failure of a crab cavity module could result in the loss of the halo in a single turn. If the halo population is too high this could cause serious damage to the collimation system or to other components of the machine. In order to fulfil all of these diagnostic requirements for halo observation in the HL-LHC, the final system should be capable of measuring halo populations at the level of $10^{-5}$ relative to that of the core.

Three techniques are currently being considered for halo monitoring in the HL-LHC:

- the use of high dynamic range cameras combined with apodization;
- core masking followed by acquisition using standard cameras;
- use of the beam gas vertex detector.

The first two techniques make use of synchrotron light and, as for the streak cameras, may require an additional light source for the final HL-LHC configuration.

### 13.7 Luminosity measurement

The measurement of the collision rate at the luminous interaction points is very important for the optimization of the machine. The LHC experiments can certainly provide accurate information about the instantaneous luminosity, but this information is often not available until stable collisions have been established, and is often missing altogether during machine development periods. For this reason simple and reliable collision rate monitors, similar to those now used in the LHC, are also needed for the HL-LHC. This measurement is currently obtained by measuring the flux of forward neutral particles generated in the collisions using fast ionization chambers installed at the point where the two beams are separated into individual vacuum chambers. The detectors (BRAN) are installed inside the neutral shower absorber (TAN) whose role is to avoid neutral collision debris and the secondary showers these induce, which reach and damage downstream machine components. The luminosity monitors therefore already operate in a very high radiation area, which for the HL-LHC is anticipated be a further ten times the nominal LHC value. For this reason the technology chosen for the HL-LHC is likely to be based on the radiation-hard LHC design [11], with the geometry adapted to the new TAN (TAXN) design. In order to further increase the radiation resistance some current features, such as their bunch-to-bunch capability, may need to be sacrificed and redundancy added.

### 13.8 Long-range beam–beam compensation

The simulated strong effect of the LHC long-range interactions inspired a proposed long-range beam–beam compensation for the LHC based on current-carrying wires [12]. In order to correct all non-linear effects the correction must be made individually in each high luminosity interaction point, with the wire generating the same integrated transverse force as the opposing beam at the parasitic long-range encounters.

The ideal location for compensation of the long-range beam–beam tune-spread is found where the beta functions are equal; there is little phase advance difference with respect to that between the long-range encounters and the IP, and where the beams are sufficiently well separated. Hence for the HL-LHC the proposed layout features compensators placed after the D1 separation dipole in IP1 and IP5 (Figure 13-6). In order to get the compensation correct for all multipoles the transverse location of a wire compensator must be on the inside of the compensated beam, i.e. between the two circulating beams, at a distance equivalent to the average long-range beam–beam separation. This poses significant technical constraints since, at the ideal longitudinal position, the transverse separation of the beams is only a few centimetres in a region with a high flux of secondary neutrons, and the transverse wire position is only a few mm from the beam. Placing and aligning a 'wire' at these locations, in particular between the two counter-rotating proton beams, will be very difficult. An alternative is to replace the 'wire' with an electron beam produced in a manner similar to that of well-established electron coolers (Figure 13-7).

Although such systems have been used for head-on beam–beam compensation at both the Tevatron (FNAL) [13] and RHIC (BNL) [14], the implementation required for the HL-LHC is beyond what has been achieved so far. Assuming an effective length of 6 m on both sides of the IP, an electron beam current of about 15–20 A would be needed, significantly higher than for previous implementations. Compensation with one system per beam in both IR1 and IR5 is also possible, and would reduce the necessary infrastructure by a factor of two, but implies electron currents of up to 40 A, almost an order of magnitude larger than what has been achieved so far.

Preliminary simulations and integration studies started to look into the feasibility of such an implementation in the LHC. However, it is clear that such infrastructure could only be installed at the same time as the upgrade of the interaction regions for the HL-LHC, ruling out the possibility of testing such a long-range beam–beam compensator beforehand. Since long-range beam–beam compensation has, so far, never been demonstrated at a long-range beam–beam limited machine, it was deemed essential to install a prototype in the LHC as soon as possible. With a test of the electron beam solution ruled out in the short term, a fall-back solution relying on wires will be pursued, with a view to installing a demonstrator in the 2015/2016 end-of-year technical stop.

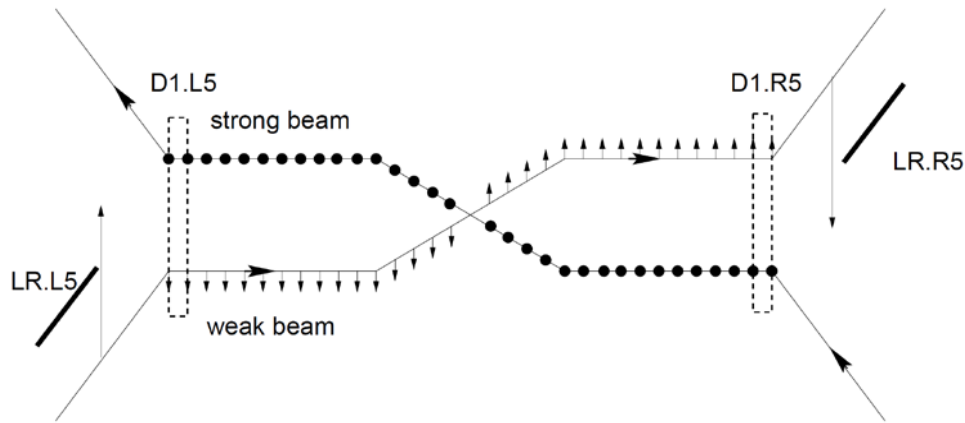

Figure 13-6: Illustration of the compensation principle [12]

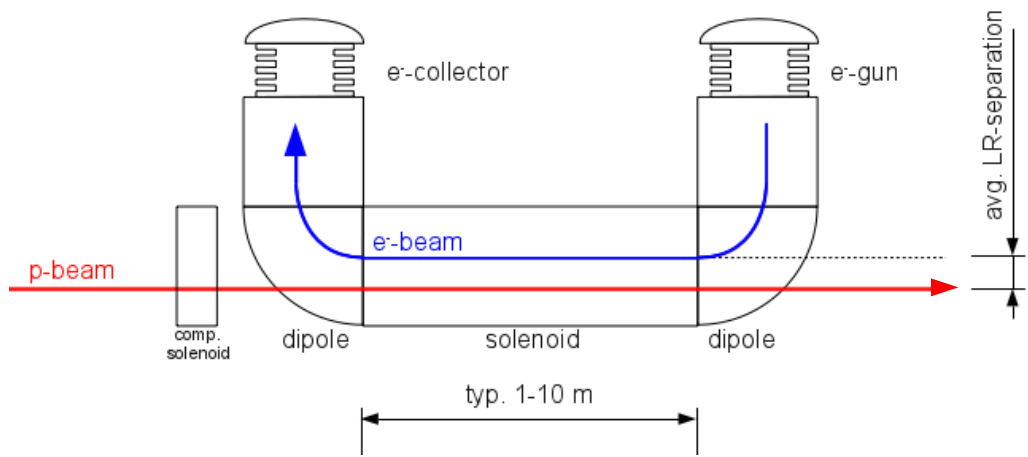

Figure 13-7: Sketch of an electron cooler type layout for long-range beam–beam compensation

### 13.8.1 Long-range beam–beam demonstrator

The only solid objects that can approach the beam accurately to within 10 $\sigma$ or less are the LHC collimators. By embedding a wire in such a collimator it would be possible to use the collimator as a host for a demonstrator version of a long-range beam–beam compensator. The best compensation effect in this scenario is obtained by a wire in the tertiary collimators (TCT) located just in front of the D2 magnet. A 1 m long wire at this location would require a DC current of some 180 A at a distance of 9.5 $\sigma$ to the beam or over 200 A at a distance of 11 $\sigma$. These values correspond to a symmetric layout with one compensator left of the IP and another on the righthand side, a set-up which will probably be necessary since the ratio of the horizontal and vertical beta functions are not equal at the TCT locations.

Integration of DC-powered wires into collimator jaws seems the only possibility to make realistic beam tests before embarking on a final implementation of the wires for LHC high luminosity operation. This integration itself requires the solution of many important technical issues:

- no interference of the wires with the nominal operation of the collimators;

- transfer of 1 kW resistive heat loss in the wire by heat conduction to the water-cooled collimator jaw;

- shielding of the wire from the beam through a thin metallic layer for impedance reasons.

The design of such a wire-in-jaw tertiary collimator is well advanced (Figure 13-8) and the production of four such collimators is foreseen to start before the end of 2014. The necessary DC cables and power converted infrastructure has been installed during LHC LS1.

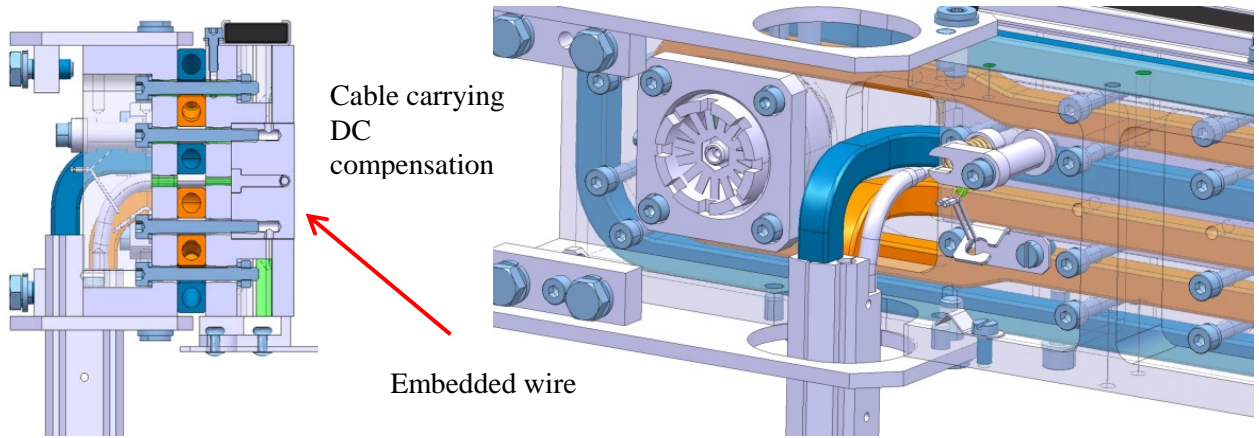

Figure 13-8: Technical drawing of the wire-in-jaw collimator design

The earliest date at which such collimators equipped with a long-range beam–beam compensation wire can be installed in the LHC is the end of year technical stop 2015/2016. Machine experiments are then planned to validate the coherence of predictions by simulation.